\catcode`\@=11

\newif\if@fewtab\@fewtabtrue
{\count255=\time\divide\count255 by 60
\xdef\hourmin{\number\count255}
\multiply\count255 by-60\advance\count255 by\time
\xdef\hourmin{\hourmin:\ifnum\count255<10 0\fi\the\count255}}
\def\ps@draft{\let\@mkboth\@gobbletwo
    \def\@oddfoot{\hbox to 7 cm{\tiny \versionno
       \hfil}\hskip -7cm\hfil\rm\thepage \hfil {\tiny\draftdate}}
    \def\@oddhead{}
    \def\@evenhead{}\let\@evenfoot\@oddfoot}
\def\draftdate{\number\month/\number\day/\number\year\ \ \ \hourmin }

\global\def\draftcontrol{0}

\def\draftcite#1{\ifnum\draftcontrol=1#1\else{}\fi}
\def\@lbibitem[#1]#2{\item{}\hskip -3\hbox to 2cm
{\hfil$\scriptstyle\draftcite{#2}$}\hskip
1cm[\@biblabel{#1}]\if@filesw
     {\def\protect##1{\string ##1\space}\immediate
      \write\@auxout{\string\bibcite{#2}{#1}}}\fi\ignorespaces}

\def\@bibitem#1{\item\hskip -3cm \hbox to 2cm
{\hfil {\footnotesize\draftcite{#1}}}\hskip 1cm
\if@filesw \immediate\write\@auxout
       {\string\bibcite{#1}{\the\value{\@listctr}}}\fi\ignorespaces}

\def\citen#1{\if@filesw \immediate\write \@auxout {\string\citation{#1}}\fi%
\@tempcntb\m@ne \let\@h@ld\relax \def\@citea{}%
\@for \@citeb:=#1\do {\@ifundefined {b@\@citeb}%
    {\@h@ld\@citea\@tempcntb\m@ne{\bf ?}%
    \@warning {Citation `\@citeb ' on page \thepage \space undefined}}%
    {\@tempcnta\@tempcntb \advance\@tempcnta\@ne
    \setbox\z@\hbox\bgroup\ifcat0\csname b@\@citeb \endcsname \relax
    \egroup \@tempcntb\number\csname b@\@citeb \endcsname \relax
    \else \egroup \@tempcntb\m@ne \fi \ifnum\@tempcnta=\@tempcntb
    \ifx\@h@ld\relax \edef \@h@ld{\@citea\csname b@\@citeb\endcsname}%
    \else \edef\@h@ld{\hbox{--}\penalty\@highpenalty
    \csname b@\@citeb\endcsname}\fi
    \else \@h@ld\@citea\csname b@\@citeb \endcsname \let\@h@ld\relax \fi}%
\def\@citea{,\penalty\@highpenalty\hskip.13em plus.13em minus.13em}}\@h@ld}
\def\@citex[#1]#2{\@cite{\citen{#2}}{#1}}%
\def\@cite#1#2{\leavevmode\unskip\ifnum\lastpenalty=\z@\penalty\@highpenalty\fi%
  \ [{\multiply\@highpenalty 3 #1%
  \if@tempswa,\penalty\@highpenalty\ #2\fi}]}   %
\makeatother 

\catcode`\@=12

\def\A             {Algebra}
\def\alg           {algebra}

\def\Bc            {Boundary condition}
\def\be            {\begin{equation}}
\def\Be            {\mbox{B$^C$}}
\def\bearl         {\begin{array}{l}}
\def\bearll        {\begin{array}{ll}}
\def\bearlll       {\begin{array}{lll}}
\def\bfe           {{\bf1}}

\def\c             {\frac1{\sqrt2}\,(1{+}\sqrt3)}
\def\cala          {{\mathfrak A}}
\def\calap         {{\bar{\mathfrak A}}}

\def\calg          {{\cal G}}
\def\calh          {{\cal H}}
\def\caln          {{\cal N}}

\def\cft           {conformal field theory}

\def\cfts          {conformal field theories}
\def\chii          {\raisebox{.15em}{$\chi$}}
\def\con           {conformal }
\def\Con           {Conformal }
\def\dl            {\mathbb }
\def\dyd           {Dynkin diagram}
\def\dz            {\frac12\,d}
\def\ee            {\end{equation}}
\def\eE            {{\rm e}}
\def\eear          {\end{array}}
\def\eq            {\,{=}\,}
\newcommand\erf[1] {(\ref{#1})}
\newcommand\Frac[2]{\mbox{$\frac{#1}{#2}$}}
\def\furu          {fusion rule}
\def\futnote#1     {\footnote{~#1}\ }
\def\Hom           {{\rm Hom}}
\newcommand\hsp[1] {\mbox{\hspace{#1 em}}}
\def\hy            {$\mbox{-\hspace{-.66 mm}-}$}

\def\ii            {{\rm i}}
\def\iN            {\,{\in}\,}
\def\irmod         {irreducible module}
\def\irrep         {irreducible representation}
\long\def\labl#1   {\label{#1}\ee \ifnum\draftcontrol=1
                   \mbox{ }\\[-12 mm]\query{#1}\\[5 mm] \fi}
\def\lambdap       {\lambda^{\!+}_{\phantom i}}

\def\Lie           {Lie group}
\def\modinv        {modular invarian}
\def\Modinv        {Modular invarian}
\def\oa            {{\rm o}}
\def\oa            {o}
\def\ob            {\check\oa}
\def\ot            {\raisebox{.07em}{$\scriptstyle\otimes$}}

\def\Rangle        {\rangle\!\rangle}
\def\rep           {representation}
\def\q             {quantum }
\def\Q             {Quantum }
\def\qft           {quantum field theory}
\long\def\query#1{\hskip 0pt{\vadjust{\everypar={}\small\vtop to 0pt{\hbox{}%
     \vskip -13pt\rlap{\hbox to 49.0pc{\hfil{\vtop{\hsize=8pc\tolerance=6000%
     \hfuzz=.5pc\rightskip=0pt plus 3em\noindent#1}}}}\vss}}}}%
\newcommand\sect[1]{\section{#1}}

\def\sa            {{\rm s}}
\def\sa            {s}
\def\sb            {\check\sa}
\def\sss           {\scriptscriptstyle}
\def\stc           {statistic}
\def\twodim        {two-dimensional}

\def\va            {{\rm v}}
\def\va            {v}
\def\vb            {\check\va}

\def\wrtt          {with respect to the }

\def\wzwm          {WZW model}

\def\wzwts         {WZW theories}
\def\zet           {{\dl Z}}

\documentclass[12pt]{article} \usepackage{amssymb,amsfonts,latexsym}
\usepackage[dvips]{graphics} \usepackage{epsfig}
\begin{document}


\begin{flushright}  {~} \\[-1cm]
{\sf hep-th/0006181}\\{\sf PAR-LPTHE 00-26}\\{\sf ESI-907}
\\[1mm]
{\sf June 2000} \end{flushright}

\begin{center} \vskip 14mm
{\Large\bf SOLITONIC SECTORS, $\alpha$-INDUCTION} \\[4mm]
{\Large\bf AND SYMMETRY BREAKING BOUNDARIES}\\[20mm]
{\large J\"urgen Fuchs$\;^1$ \ and \ Christoph Schweigert$\;^2$}
\\[8mm]
$^1\;$ Institutionen f\"or fysik~~~~{}\\
Universitetsgatan 1\\ S\,--\,651\,88\, Karlstad\\[5mm]
$^2\;$ LPTHE, Universit\'e Paris VI~~~{}\\
4 place Jussieu\\ F\,--\,75\,252\, Paris\, Cedex 05
\end{center}
\vskip 18mm
\begin{quote}{\bf Abstract}\\[1mm]
We develop a systematic approach to boundary conditions 
that break bulk symmetries in a general way such that left and right 
movers are not necessarily connected by an automorphism. In the
context of string compactifications, such boundary conditions typically
include non-BPS branes.\\
Our formalism is based on two dual fusion rings, one for the bulk and one 
for the boundary fields. Only in the Cardy case these two structures coincide.
In general they are related by a version of $\alpha$-induction.
Symmetry breaking boundary conditions correspond to solitonic sectors.
In examples, we compute the annulus amplitudes and boundary states.
\end{quote}
\newpage


\sect{Introduction}

Conformally invariant boundary conditions of two-dimensional conformal field 
theories arise in the study of defects in systems of condensed matter physics, 
of percolation probabilities and of (open) string perturbation theory in 
the background of D-branes. They are presently under active investigation.
Boundary conditions that preserve all bulk symmetries 
for theories with charge conjugation modular invariant
have been treated by Cardy \cite{card9}. The two basic results
are the following: Boundary conditions are labelled by the primary
fields of the theory, and the annulus multiplicities are given by the fusion
rules. Together with the information that all bulk symmetries are preserved, 
these two results allow in particular to recover the so-called boundary states,
which encode all one-point amplitudes on the disk.

More recently, these results have been generalized in several directions
\cite{fuSc10,fuSc11,fuSc12}. In particular, those boundary conditions
have been classified which preserve an abelian orbifold subalgebra of the algebra
$\cala$ of bulk symmetries, i.e.\ for which the preserved symmetries can be 
characterized as the subalgebra $\calap\eq\cala^G$ of symmetries that are 
fixed by some abelian group $G$ of
automorphisms of $\cala$. Boundary states for such boundary conditions have
been given explicitly, and the integrality of the annulus coefficients was
proven. It was also shown that correlation functions in the presence of
such boundary conditions can be written as linear combinations of twisted conformal
blocks. As a special case, the boundary states can be expressed in terms
of twisted Ishibashi states $|\lambda,\omega \Rangle$, which are characterized 
by the identity
  \be  \left(Y_n \ot \bfe + (-1)^{\Delta_Y-1} \bfe\ot \omega(Y_{-n}) \right) 
  |\lambda,\omega \Rangle = 0 \labl1
for every primary
field $Y$ (of conformal weight $\Delta_Y$) in the chiral symmetry algebra
$\cala$. Here $\omega\iN G$ is an automorphism of the bulk symmetries 
that leaves the Virasoro algebra invariant, $\omega(L_n)\eq L_n$.
In a sense, the relations \erf1 express the fact that at the boundary
left movers and right movers are connected by the automorphism $\omega$.
The automorphism $\omega$ has been called 
the automorphism type, or gluing automorphism, of the
boundary condition. We will say that such boundary conditions possess a
definite automorphism type, in this case $\omega$.

In the present letter, we study more general patterns of symmetry
breaking by boundaries, in which left movers and right movers are not
necessarily related any more by automorphisms. In more precise terms, this means
that the boundary conditions preserve a subalgebra $\bar\cala$ of
the algebra $\cala$ of bulk symmetries that cannot any longer be characterized
as a fixed algebra under some group of automorphisms. We refer to such 
boundary conditions as boundary conditions {\em without automorphism type\/},
or without gluing automorphism.
Examples of such boundary conditions appear already for the $\zet_2$-orbifold 
of a compactified free boson. Other examples are provided by various conformal 
embeddings; boundary conditions associated to conformal embeddings
have been studied in \cite{bppz,bppz2}, in particular in their relation with
certain graph algebras \cite{pasq4,dizu1,difr2,pezu2}.

Boundary conditions without automorphism type are of direct relevance in
string theory: they correspond to non-BPS branes.
Indeed, every chiral algebra automorphism $\omega$ maps the vertex operator of 
a space-time supercharge to the vertex operator of another supercharge. Therefore
validity of \erf1 immediately
implies that boundary conditions with automorphism type preserve half of
the space-time supersymmetries, and hence are BPS. 

The purpose of the present note is to generalize Cardy's results and those
of \cite{fuSc10,fuSc11,fuSc12} once more. We consider conformally
invariant boundary conditions of a rational conformal field theory
with chiral algebra $\cala$ that preserve a subalgebra $\calap$ of $\cala$
that is still rational, but otherwise arbitrary. 
We give a natural labelling of the boundary conditions
and compute the annulus coefficients. By a modular transformation, this
allows to determine the boundary states. 

For certain conformal field theories, one can construct (nets of) factors; 
their irreducible local sectors (inner unitary equivalence class of \rep s)
are in one-to-one correspondence to the primary fields of the \cft. 
Our main tool is an adaptation of a certain form of induction for sectors, 
the so-called $\alpha$-{\em induction\/}, which was developped
\cite{lore,xu3,boev123,boek2,boek3} in the framework of subfactor theory. Applying
$\alpha$-induction to a sector of the subfactor, it produces a sector of the 
ambient factor. Among the sectors obtained this way there are ordinary, 
local, sectors as well as solitonic sectors.  

For the purposes of this letter, we do not have to know the 
relevant nets of subfactors and their sectors in any detail. Rather, we
simply {\em postulate\/} that the process of $\alpha$-induction works at the
level of the representation category of the conformal field theory under
investigation. Thus we regard the irreducible sectors as the primary fields, 
or rather as the associated basis elements of the fusion ring;
general sectors correspond to arbitrary elements of the fusion ring, and the
composition of sectors is simply the fusion product. In fact, all we need to
know is the action of $\alpha$-induction on primary fields. It will provide 
us with solitonic sectors which precisely label symmetry breaking boundary
conditions. Moreover, the fusion of these sectors
will provide us with the annulus multiplicities.
As in the Cardy case, these data, together with the preserved
symmetries $\calap$, allow to construct the boundary states.

In section 2 we discuss the labelling problem for bulk and boundary fields.
Motivated by constructions from topological field theory, we are led
to the concepts of bulk and boundary categories. Our
prescription for the boundary category is presented in section 3.
It is based on imposing a version of $\alpha$-induction at the level
of the fusion rules. In section 4 two illustrative examples are analyzed.

\section{Symmetry breaking boundary conditions}

Before we explain the case of boundary conditions without automorphism
type, we briefly rephrase some of the results of \cite{fuSc11,fuSc12} on
boundary conditions that do possess an automorphism type.
As was shown there, boundary conditions leaving $\calap\eq\cala^G$ invariant
correspond to orbits $[\bar\mu,\psi]$
of primary fields $\bar\mu$ of the $\calap$-theory with respect to a group 
$\calg\,{\cong}\,G^*$ of simple current fields (the degeneracy label
$\psi$ is a character of a suitable subgroup of $\calg$).
The monodromy charge \cite{scya6} of $\bar\mu$ with respect to $\calg$
is not restricted. These labels $[\bar\mu,\psi]$, in turn, can be seen to
correspond to representations of $\cala$; these are {\em twisted
representations} when the monodromy charge of $\bar\mu$ is not zero.\,%
 \futnote{When the action of the twists is only projective, additional
 subtleties arise.}
Twisted \rep s of vertex operator algebras have been investigated e.g.\ in
\cite{lepo4,FRlm,dong2,dolm3}.
The notion of fusion of such representations has also been studied 
\cite{gabe7}, but little is known about the resulting fusion ring. 
However, one can re-write the annulus amplitudes derived in \cite{fuSc11}
as sums of characters of the twisted representations. It is therefore 
reasonable to expect that the annulus coefficients determined in \cite{fuSc11} 
precisely coincide with the fusion rules of the twisted representations.

A second ingredient we will need is the description of correlation functions in
the presence of boundaries through three-dimensional topological field
theory that was developped in \cite{fffs2,fffs3} for symmetry preserving
boundary conditions. In that context, a three-dimensional manifold $M_X$,
the connecting manifold, was constructed to compute correlators on a 
two-dimensional world sheet $X$. $M_X$ has a
boundary, and this boundary is isomorphic to the so-called double 
$\hat X$ of $X$. The connecting manifold is universal in the sense that
it is the same for {\em all\/} rational \cfts. One also needs to prescribe
a Wilson graph in $M_X$. Bulk points on the world sheet $X$ possess two
pre-images on its double $\hat X\eq \partial M_X$, and these two points 
are connected by a natural interval in $M_X$. A Wilson line carrying
the bulk label is placed in that interval. 

As for the boundary data, a circular Wilson line must be placed parallel to 
each boundary component. Insertions of boundary fields are linked with short
Wilson lines to the corresponding circular Wilson line.
This is summarized in the following picture for the case when $X$ is the
disk (then $\hat X$ is the sphere and $M_X$ is a solid three-ball):
  \be
  \raisebox{-3.4em}{\scalebox{0.2}{\includegraphics{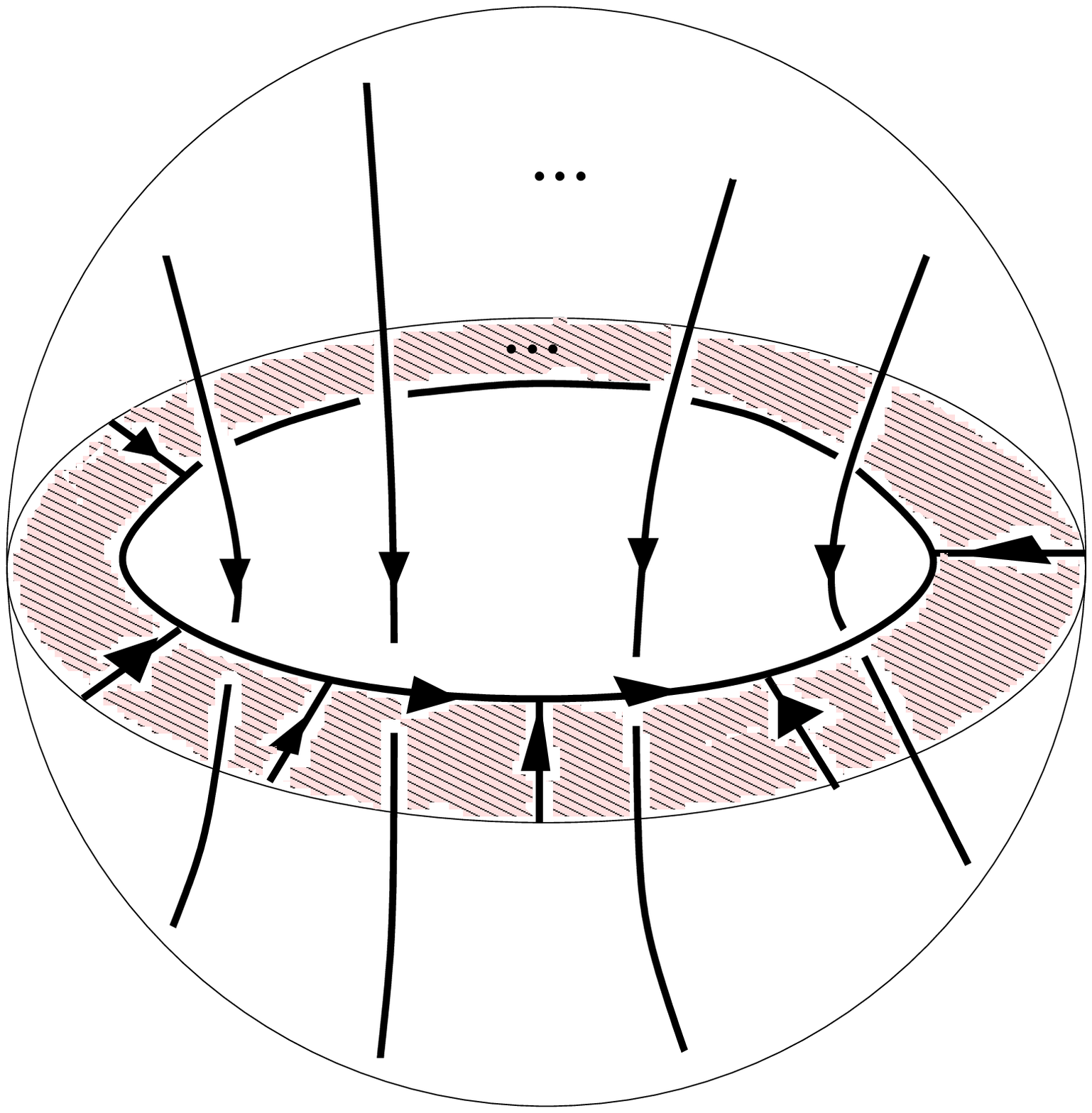}}}
  \ee
We will assume that the Wilson graph is
universal as well, i.e.\ that the same graph is still to be used for symmetry 
breaking boundary conditions. The boundary graph must be labelled, too.
In the Cardy case, we use the same type of labels as for the bulk components,
i.e.\ primary fields of $\cala$. This nicely fits together with Cardy's
result that symmetry preserving boundary conditions are in one-to-one 
correspondence with primary fields. Moreover, in this description trivalent
vertices, couplings, appear naturally in the boundary graph. They involve 
two boundary conditions and the chiral label of a boundary insertion.

We remark that typically the Wilson graph is {\em not\/} connected. What
is crucial in the present context is that there is never a Wilson line 
that connects a bulk insertion with a boundary insertion or with a segment
of the Wilson graph that encodes a boundary condition.
We are thereby led to the following general picture. Boundary conditions
as well as field insertions on the boundary should be characterized by 
basis elements of the same `fusion ring', and the structure constants of this
fusion ring should coincide with the annulus coefficients. In more technical
terms, the labels should correspond to the (isomorphism classes of)
simple objects of a suitable tensor category.
We call this structure the {\em boundary category\/}. Bulk fields, on the other
hand, will have to be described by a different structure, namely as the simple 
objects of a {\em bulk category\/}; their treatment is beyond the scope of the 
present note. Notice that the interpretation in terms of boundary insertions 
requires that associated to each boundary label there comes a 
natural state space, which is a module over (at least) the Virasoro algebra.

In Cardy's case -- torus partition function of charge conjugation type, 
and only symmetry preserving boundary conditions -- both the boundary
and the bulk category are just the one associated to the fusion ring of the 
full bulk symmetry $\cala$, and the state space associated to a label $\lambda$ 
is nothing but the corresponding irreducible $\cala$-module $\calh_\lambda$. 
For a general boundary insertion, the state space must be 
an appropriate generalization of a twisted representation, or in other words,
a {\em solitonic\/} \cite{froh3,fred10,muge5} sector. In Cardy's case, the 
tensor category has actually a lot more structure. In particular it is 
{\em modular\/} \cite{TUra}, i.e., roughly speaking, there is 
a braiding \cite{frrs,frga}, leading to the unitarity of the modular matrix $S$, 
as well as a twist of Wilson lines, which corresponds to the modular matrix $T$.

This is too much structure than may be expected to be present for a boundary 
category in the general case. As examples of twisted representations show, the 
twist of a Wilson line will, in general, no longer be a well-defined 
operation on the boundary. Correspondingly, the boundary sectors
will not have a unique conformal weight up to integers. (The bulk category,
on the other hand, will still possess a $T$-matrix.) Similarly, neither the bulk 
nor the boundary category will possess a braiding any more, in general. However,
it is still possible to braid an object of the bulk category with an
object of the boundary category. This braiding allows us to establish
a generalization of the diagonalizing matrix $\tilde S$ that was introduced in
\cite{fuSc11}. We expect that this matrix $\tilde S$ is square; nevertheless
its two indices take values in two different sets: the rows are labelled by the
bulk fields, whereas the columns are labelled by the boundary conditions.
Up to normalization, $\tilde S$ provides the
coefficients for the expansion of the boundary states \wrtt boundary conformal
blocks (generalized Ishibashi states). We conjecture that the matrix $\tilde S$ 
obtained this way is invertible and hence in particular indeed a square matrix
(and can be chosen unitary). This implies that the bulk and the boundary 
category have the same number of simple objects; as a consequence, the number of 
boundary conditions can be read off from the bulk modular invariant. 
The equality between the number of bulk fields and the number of boundary
conditions has been derived in \cite{fuSc11} for boundary conditions with 
definite automorphism type; we will see in the examples below that it holds 
for more general symmetry breakings as well.

\sect{Solitonic sectors from $\alpha$-induction}

According to the reasoning above, we would like to view the boundary labels
$\mu$ as simple objects of a suitable category, and accordingly the associated 
intertwiners as morphisms of that category. It will, however, be 
important that we can regard the sectors also as (isomorphism 
classes of) \rep s -- including both ordinary and solitonic \rep s -- of the 
chiral algebra or vertex operator algebra $\cala$. These representation
spaces provide the spaces of open string states whose partition
function is the annulus amplitude.

How can we then obtain the boundary category? We wish to find `solitonic'
$\cala$-repre\-sen\-ta\-tions. Fortunately, there is one situation in
which this can be done totally explicitly. Namely, chiral \wzwts\ can also
be analyzed in the framework \cite{HAag} of nets of operator algebras. In that
context, one can employ the notion of $\alpha$-induction to arrive at
solitonic sectors of the chiral theories \cite{lore,boev123}.

The concrete construction of operator algebras for more general chiral \cfts\ 
is a difficult problem; see \cite{gafr,brgl}, as well as \cite{xu6,xu10} for 
the case of coset and orbifold theories. For the purposes of the present note,
we need not address these questions which are definitely quite
important. Rather, we will only abstract from $\alpha$-induction and its 
adjoint operation, $\sigma$-restriction, a few properties at the level of
fusion rings. We present these properties in the form of a recipe.
(But we expect that they are indeed realized in any decent \cft, and
actually that the existence of such an induction procedure can be
entirely established in the context of the relevant tensor categories.)

We start by prescribing those bulk symmetries that are preserved by the
boundary conditions. These symmetries must form a consistent rational 
subalgebra $\calap$ of the chiral algebra $\cala$. We assume that the collection 
of all (isomorphism classes of) ordinary -- i.e.\ non-solitonic --
\irmod s $\bar\calh_{\bar\lambda}$ over $\calap$ gives rise to a modular 
fusion ring with basis $\{\bar\Phi_{\bar\lambda}\}$. From the embedding of 
$\calap$ into $\cala$ we determine how the vacuum module $\calh_\Omega$ of $\cala$
decomposes into a direct sum of irreducible $\calap$-modules:
  \be \calh_\Omega = \bigoplus_{\bar\mu} b_{\bar\mu}\, \bar\calh_{\bar\mu}
  \,.  \ee
(Thus $b_{\bar\lambda}\iN\zet_{\ge0}$ is the multiplicity with which
the $\calap$-module $\bar\calh_{\bar\lambda}$ appears in the vacuum module of
the $\cala$-theory.)
This allows us to introduce the element 
  \be  \bar\theta := \sum_{\bar\mu} b_{\bar\mu}\, \bar\Phi_{\bar\mu} \ee
of the fusion ring of $\calap$; we refer to $\bar\theta$ as the {\em extending
sector\/} of the $\calap$-theory.

We now construct a new fusion ring as follows. First, it is generated by objects 
$\alpha_{\bar\lambda}$ for each basis element $\bar\Phi_{\bar\lambda}$ of the fusion
ring of $\calap$. The fusion product is defined by
  \be  \alpha_{\bar\lambda} \star \alpha_{\bar\mu}
  := \alpha_{\bar\lambda\star\bar\mu} \,,  \labl4
and we also require that $\alpha_{\bar\lambda\oplus\bar\mu}\eq\alpha_{\bar\lambda}
\,{+}\,\alpha_{\bar\mu}$ and $\alpha_{\bar\lambdap}\eq(\alpha_{\bar\lambda})^+_{}$.
This would not constitute anything new beyond what is encoded in
the fusion ring of $\calap$, 
were it not for another piece of information. Namely, the fusion ring element
$\alpha_{\bar\lambda}$ is also supposed to represent a -- possibly twisted or 
solitonic -- representation of $\cala$, which for brevity we denote by the same 
symbol. An important point is that even for irreducible $\bar\lambda$ the 
$\cala$-representation $\alpha_{\bar\lambda}$ need not necessarily be irreducible, 
and that the $\alpha_{\bar\mu}$, respectively their irreducible sub\rep s, 
for different values of $\bar\mu$ are allowed to be isomorphic. (Thus in particular 
the $\alpha_{\bar\lambda}$ generically do not form a basis of the fusion ring.)
In view of Schur's lemma, this information is conveniently encoded in the 
intertwiner spaces $\Hom_\cala(\alpha_{\bar\lambda},\alpha_{\bar\mu})$. 
For instance, $\alpha_{\bar\lambda}$ is irreducible if and only if
$\Hom_\cala(\alpha_{\bar\lambda},\alpha_{\bar\lambda})$ is one-dimensional. Also, 
when $\alpha$ is a simple and $\beta$ any arbitrary object of the fusion category,
then the dimension of $\Hom_\cala(\alpha,\beta)$ tells us how many times $\alpha$
appears in the decomposition of $\beta$. The $\Hom$ spaces are defined in terms 
of the intertwiner spaces in the fusion category of $\calap$ as follows:
  \be \Hom_\cala(\alpha_{\bar\lambda},\alpha_{\bar\mu})
  := \Hom_{\calap}(\bar\Phi_{\bar\lambda}, \bar\theta\star\bar\Phi_{\bar\mu}) \,.
  \labl2

This system of $\Hom$ spaces obeys tight consistency constraints. For example,
from $\Hom_\cala(\alpha_{\bar\lambda},\linebreak[0]\alpha_{\bar\lambda})$ we 
compute the number $n_{\bar\lambda}$ of irreducible subsectors of 
$\alpha_{\bar\lambda}$. If one would just prescribe an extending sector 
$\bar\theta$ at random, one might find contradictions of the type that more 
than $n_{\bar\lambda}$ irreducible sectors have non-trivial intertwiners with 
$\alpha_{\bar\lambda}$. The existence of a system of $\Hom$ spaces that is
free of contradiction is therefore highly non-trivial and requires
special properties of $\bar\theta$. A necessary condition is of course that 
all irreducible $\calap$-subsectors of $\bar\theta$ are mutually local, but 
this condition is typically far from being sufficient. It would be rewarding 
to find a characterization of consistent extending sectors purely at the level 
of fusion rings. It will then be particularly interesting to compare the
problem of classifying consistent extending sectors with the problem
of classifying modular invariant partition functions of extension type.

It is also important that along with $\alpha$-induction there comes an 
``adjoint'' operation, known as $\sigma$-{\em restriction\/}. Namely, every
sector $\beta$ of $\cala$, whether solitonic or not, may be seen as a
(typically reducible) sector of $\calap$, which we denote as
$\sigma(\beta)$. Induction and restriction are related by the reciprocity 
relation
  \be \Hom_{\cala}(\alpha_{\bar\lambda},\beta) \,\cong\, 
  \Hom_{\calap}(\bar\lambda, \sigma(\beta)) \,.  \labl3
This implies that
  \be  \sigma(\alpha_{\bar\mu}) = \bar\theta \star \bar\mu  \ee
and allows us to decompose induced solitonic sectors
into irreducible $\calap$-sectors.

\smallskip

Let us pause and compare these ideas to the situation studied in \cite{fuSc11}. 
In that case, $\bar\theta$ can be written as a sum
over so-called simple current sectors $\bar J$ which form a finite abelian group
$\calg$ under fusion, and each such simple current appears with multiplicity one: 
  \be  \bar\theta = \sum_{\bar J\in\calg} \bar J \,.  \ee
Formula \erf2 then just summarizes how the fusion rules of a simple current 
extension are related to those of the original theory (in the category theoretical 
setting, this is discussed in \cite{brug2,muge6,sawi2}). However, it only allows 
for a direct determination of the extended fusion rules as long as no fixed points
-- that is, sectors $\bar\lambda$ with $\bar J\,{\star}\,\bar\lambda\eq
\bar\lambda$ for some $\bar J\iN\calg$ -- are involved. Indeed, in the simple 
current situation  the induced sector $\alpha_{\bar\lambda}$ is reducible if 
and only if 
$\bar\lambda$ is a fixed point. The decomposition of $\alpha_{\bar\lambda}$ for 
a fixed point is precisely what is known as \cite{scya6,fusS6} fixed point 
resolution in the theory of simple current extensions.

In the general case there is the following analogue of the problem caused
by simple current fixed points. It can happen that 
the relations \erf2 do not provide enough information for decomposing all
$\alpha_{\bar\lambda}$ into irreducible sectors. 
In that case, the category must be enlarged: sufficiently
many additional irreducibles have to be introduced to provide subobjects.
There exists a general procedure for doing so \cite{loro,muge6}. But
unfortunately fully explicit formulae, in particular for the modular
$S$-matrix of the enlarged theory, are only known in the simple current
case \cite{fusS6}, where it leads in particular to the group character 
$\psi$ that appears in the description of boundary conditions with definite 
automorphism type, see above. 

A more explicit understanding of these new 
irreducibles in the general case and in particular what their braiding 
properties with bulk fields are, might be called the {\em generalized fixed 
point problem\/}.  To be precise, the task is to express the fusion products 
of the new irreducibles in terms of chiral data of the $\calap$-theory, like 
e.g.\ the modular matrices for one-point conformal blocks on the torus. 
We consider this to be a central problem in the study of 
solitonic sectors, and hence of conformally invariant boundary conditions.\,%
 \futnote{A special version of the fixed point problem arises already
 when one aims to express the modular $S$-matrix of the $\cala$-theory
 through chiral data of the $\calap$-theory. For exceptional extensions,
 no general solution to this problem is known.}
In the present letter, we restrict
ourselves to examples where either this problem does not occur at all or where 
it can be resolved by using the knowledge about the simple current case.

Our prescription provides us explicitly with labels for the boundary
conditions and the boundary insertions. The annulus multiplicities are
just the tensor product multiplicities in the boundary category, and the
open string states are organized in terms of the induced 
sectors. The induced sectors come in two classes: ordinary, non-solitonic
sectors correspond to symmetry preserving boundary conditions, while the 
solitonic sectors are in correspondence with symmetry breaking boundary 
conditions.  In the case of boundary conditions with automorphism type, the 
latter are just the orbits with non-vanishing monodromy charge. In the subfactor 
framework, ordinary and solitonic sectors can be distinguished by their 
localization properties.

Before we support our findings by examples, we wish to add a speculative 
comment. In the operator-algebraic definition of $\alpha$-induction,
a braiding among $\calap$-sectors enters. In two dimensions
there are two independent braidings -- `over'- and `under'-braiding --
which are each others' inverse. As a consequence, there are in fact 
{\em two\/} $\alpha$-inductions, called $\alpha^\pm$. It has been shown in
\cite{boev123} that $\alpha^\pm_{\bar\lambda}$ is not solitonic if and only
if $\alpha^+_{\bar\lambda}$ and $\alpha^-_{\bar\lambda}$ are isomorphic.
This suggests that solitonic representations, and thus symmetry breaking
boundary conditions, actually come in pairs. However, only one version of 
$\alpha$-induction may be used at a time; so there is a twofold choice on which 
set of (symmetry breaking) boundary conditions one must take.
It will be interesting to see whether this can explain the observations
in \cite{husc}, where {\em two\/} distinct sets of symmetry breaking boundary
conditions were found; any two boundary conditions of the same set are 
compatible, while two boundary conditions belonging to distinct sets
are mutually incompatible.

\section{Examples}

Our general ideas are easily illustrated by examples; we present two of them. The
first example is the $E_6$-type modular invariant of $A_1$ at level $10$, which
has already been discussed extensively elsewhere \cite{prss,xu3,boev123,bppz2}.
We will show how the structures developed above allow to rederive and 
systematize the results of \cite{bppz2} on the boundary conditions of this
theory. The second example deals with the exceptional modular invariant of
$G_2$ at level 3 and is, to the best of our knowledge, new.

\smallskip

The fusion ring of $A_1$ at level $10$ has eleven simple sectors, which
we label by $\bar\mu\eq0,1,...\,,10$. In this notation, the $E_6$-type 
modular invariant of $A_1$ reads
  \be Z = |\chii_0+\chii_6|^2 + |\chii_4+\chii_{10}|^2 +
  |\chii_3+\chii_7|^2 \, . \labl z
It corresponds to the conformal embedding into $B_2$ at level $1$.
The first block comes from the vacuum $o$, the second from the vector
$v$ and the third block from the spinor $s$ of $B_2$.  The relevant aspects 
of $\alpha$-induction for this example can be found in 
\cite[Sec.\,2.2\,of\,II]{boev123}; here we summarize the most important features.

{}From the modular invariant \erf z we read off the extending sector as
$\bar\theta\eq\bar\Phi_0\,{+}\,\bar\Phi_6$. The dimensions of the $\Hom$ 
spaces are thus given by
  \be \bearl
  \dim\Hom_\cala(\alpha_{\bar\mu_1},\alpha_{\bar\mu_2})
  = \dim\Hom_\calap(\bar\Phi_{\bar\mu_1}, \bar\Phi_{\bar\mu_2}) 
  + \dim\Hom_\calap(\bar\Phi_{\bar\mu_1}, \bar\Phi_6{\star}\bar\Phi_{\bar\mu_2}) 
  \\[2.5mm] \hsp{22.7} = \delta_{\bar\mu_1,\bar\mu_2} +
  \bar\caln_{6,\bar\mu_2}^{\;\ \bar\mu_1} \,,\eear \ee
where $\bar\caln$ are the fusion rules of $A_1$ at level $10$. 
Applying this to the case $\bar\lambda_1=\bar\lambda_2$, we find that
the sectors $\alpha_{\bar\lambda}$ are irreducible for
$\bar\lambda\eq0,1,2,8,9,10$, and contain two irreducible subsectors else. 

Computing the $\Hom$ spaces between the irreducible 
$\alpha_{\bar\lambda}$ shows that they all vanish, except for
$\Hom(\alpha_2,\alpha_8)$, which is one-dimensional. Hence
the two irreducible sectors $\alpha_2$ and $\alpha_8$ are isomorphic,
$\alpha_2\,{\cong}\,\alpha_8$. Furthermore,
  \be \dim\Hom(\alpha_2,\alpha_4) = 1 = \dim\Hom(\alpha_{10},\alpha_4) \,,
  \ee
so that $\alpha_4 \,{\cong}\, \alpha_2\,{+}\,\alpha_{10}$. Similarly,
one finds $\alpha_5\,{\cong}\, \alpha_1\,{+}\,\alpha_9$ and 
$\alpha_6\,{\cong}\,\alpha_0\,{+}\,\alpha_2$.
Thus these sectors do not give rise to new irreducible sectors. According
to our general conjecture, and in accordance with the results of \cite{bppz2}, 
we expect in total 6 boundary conditions and thus one additional simple object,
which we call $\alpha^{\sss(1)}_3$. 
Indeed we have\, $\dim\Hom_\cala(\alpha_3,\alpha_{\bar\mu})\eq\delta_{\bar\mu,3}
\,{+}\dim\Hom_\calap(\bar\Phi_{\bar\mu},\bar\Phi_6{\star}\bar\Phi_3)
\eq\delta_{\bar\mu,3}\linebreak[0]{+}\dim\Hom_\calap
(\bar\Phi_{\bar\mu}, \bar\Phi_3{+}\bar\Phi_5{+}\bar\Phi_7 {+}\bar\Phi_9)$. 
So $\alpha^{\sss(1)}_3$ appears in the decompositions
  \be \alpha_3\cong \alpha_9 + \alpha^{\sss(1)}_3 \qquad
  \alpha_7\cong  \alpha_1 + \alpha^{\sss(1)}_3 \, . \ee

The $\sigma$-restriction is found from formula \erf3. First,
  \be 
  \sigma(\alpha_0)  \cong  0 \oplus 6 \,, \quad
  \sigma(\alpha_{10})  \cong  4 \oplus 10 \,, \quad
  \sigma(\alpha_3^{\sss(1)}) \cong 3\oplus 7 \,,  \ee
showing that these sectors are the three non-solitonic sectors of $B_2$
that can already be inferred from the partition function \erf z, namely
  \be  o = \alpha_0\,,\quad v = \alpha_{10}\,,\quad s = \alpha_3^{\sss(1)} 
  \,.  \ee
It is convenient to introduce a similar notation $\ob,\,\vb,\,\sb$ for
the three solitonic $B_2$-sectors; they restrict as follows:
  \be \bearl
  \sigma(\ob)\equiv\sigma(\alpha_1) \cong 1\oplus 5\oplus 7  \,,  \qquad
  \sigma(\vb)\equiv\sigma(\alpha_9) \cong 3\oplus 5\oplus 9  \,,  \\[2mm]
  \sigma(\sb)\equiv\sigma(\alpha_2) \cong 2\oplus 4\oplus 6\oplus 8
  \,. \eear \ee
It is readily checked that all annulus amplitudes reported in \cite{bppz2} can 
indeed be written as linear combinations of the corresponding six specific sums 
of $A_1$-characters.

The fusion products of the sectors $\alpha_{\bar\mu}$ are computed
with formula \erf4. For $\oa,\,\va$ and $\sa$ we get
the usual Ising fusion rules; they indeed provide the
annuli of the Cardy boundary conditions. $\alpha_0$ acts generally
as the identity under fusion. The remaining fusion rules between ordinary and
solitonic sectors turn out to be
  \be \begin{array}{lll}
  \va \star \ob = \vb \,,\ &  \va \star \vb = \ob \,, &  
  \va \star \sb = \sb \,, \\[1.5mm]
  \sa \star \ob = \sb \,,  &  \sa \star \vb = \sb \,, &  
  \sa \star \sb = \ob + \vb \,.  \end{array}\ee
The fusion between two solitonic sectors produces ordinary as well as solitonic
sectors; we find
  \be \begin{array}{ll}
  \ob \star \ob = \oa + \sb \,, &  \vb \star \vb = \oa + \sb \,, \\[1mm]
  \ob \star \vb = \va + \sb \,, &  \vb \star \sb = \sa + \ob + \vb \,, \\[1mm]
  \ob \star \sb = \sa + \ob + \vb \,,\ &  \sb \star \sb = \oa + \va + 2\sb \,. 
  \end{array}\ee
These fusion products exactly give the annulus multiplicities that have been 
found by different arguments in \cite{bppz2}.
Also the $\tilde S$ matrix can be computed. It reads
  \be  \tilde S = \frac1d \, \left( \begin{array}{cccccc}
  1    &  \sqrt2      &  1    &  \c   &  1{+}\sqrt3  &  \c   \\[.8mm]
  \dz  &  0           & -\dz  &  \dz  &  0           & -\dz  \\[.8mm]
  1    &  -\sqrt2     &  1    & -\c   &  1{+}\sqrt3  & -\c   \\[.8mm]
  \c   &  1{+}\sqrt3  &  \c   & -1    &  -\sqrt2     & -1    \\[.8mm]
  \dz  &  0           & -\dz  & -\dz  &  0           &  \dz  \\[.8mm]
  \c   & -1{-}\sqrt3  &  \c   &  1    &  -\sqrt2     &  1    
  \eear \right)  \ee
with $d:=1/2\sqrt{3+\sqrt3}$.

\smallskip
Our second example is the exceptional modular invariant of $G_2$ at
level 3. It reads
  \be Z = | \chii_{00} + \chii_{11}|^2 + 2\, |\chii_{02}|^2 \ee
and describes the conformal embedding into $E_6$ at level 1. Here 
$G_2$-sectors are characterized by their highest weights. The multiplicity 
two in the second term of $Z$ indicates that the 27-dimensional \irrep\ 
of $E_6$ and its conjugate restrict to the same irreducible $G_2$-\rep.
$G_2$ at level $3$ has six irreducible sectors. A careful analysis of
the $\Hom$ spaces shows that $\alpha_{01}, \alpha_{03}$ and $\alpha_{10}$
are irreducible and all isomorphic. $\alpha_{00}$ is, as always, irreducible,
and indeed not isomorphic to $\alpha_{01}$. $\alpha_{11}$ contains
two irreducibles, and one finds $\alpha_{11}\,{\cong}\,\alpha_{00}+
\alpha_{01}$ so that it does not give rise to any new irreducibles.
Finally, $\dim\Hom(\alpha_{02},\alpha_{02})\eq3$, and $\alpha_{02}$
contains $\alpha_{01}$ as a subobject. We choose the notation
$\alpha_{02}^{\sss(\pm)}$ for its two other subobjects:
  \be  \alpha_{02} = \alpha_{01} + \alpha_{02}^{\sss(+)}
  +\alpha_{02}^{\sss(-)} \, . \ee 

The computation of the $\sigma$-restriction is straightforward, too.
We get
  \be\begin{array}{l}
  \sigma(\alpha_{01}) = \bar\Phi_{01} + \bar\Phi_{02}
    + \bar\Phi_{03} +\bar\Phi_{10}+\bar\Phi_{11} \,,  \\[2mm]
  \sigma(\alpha_{00}) = \bar\Phi_{00} + \bar\Phi_{11} \,,\qquad
  \sigma(\alpha_{02}) = \sigma(\alpha_{01})  + 2\, \bar\Phi_{02} 
  \,, \end{array}\ee
from which we also learn that $\sigma(\alpha_{02}^{\sss(\pm)})\eq\bar\Phi_{02}$.
We can therefore identify $\alpha_{00}$ as the vacuum sector of the 
$E_6$-theory and $\alpha_{02}^{\sss(\pm)}$ as the sectors corresponding to the
two 27-dimensional \irrep s of $E_6$. In addition there is a single solitonic 
sector, given by $\alpha_{01}$. Notice that we obtain again the same number of
simple objects in the bulk and in the boundary category.

It is readily checked that the fusion rules of $\alpha_{00}$ and
$\alpha_{02}^{\sss(\pm)}$ are indeed the $\zet_3$ fusion rules of $E_6$ at
level 1. The fusion rules involving $\alpha_{01}$ turn out to be
  \be\begin{array}l
  \alpha_{01}\star \alpha_{00} = \alpha_{01} \,, \qquad\
  \alpha_{01}\star \alpha_{02}^{\sss(+)}
  = \alpha_{01} = \alpha_{01}\star \alpha_{02}^{\sss(-)} \,, \\[2mm]
  \alpha_{01}\star \alpha_{01} = \alpha_{00} + 3\, \alpha_{01}
  + \alpha_{02}^{\sss(+)} + \alpha_{02}^{\sss(-)}
  \,.  \end{array}\ee
Thus the fusion graph of $\alpha_{01}$ looks like
  \be
  \raisebox{-2.8em}{\scalebox{0.14}{\includegraphics{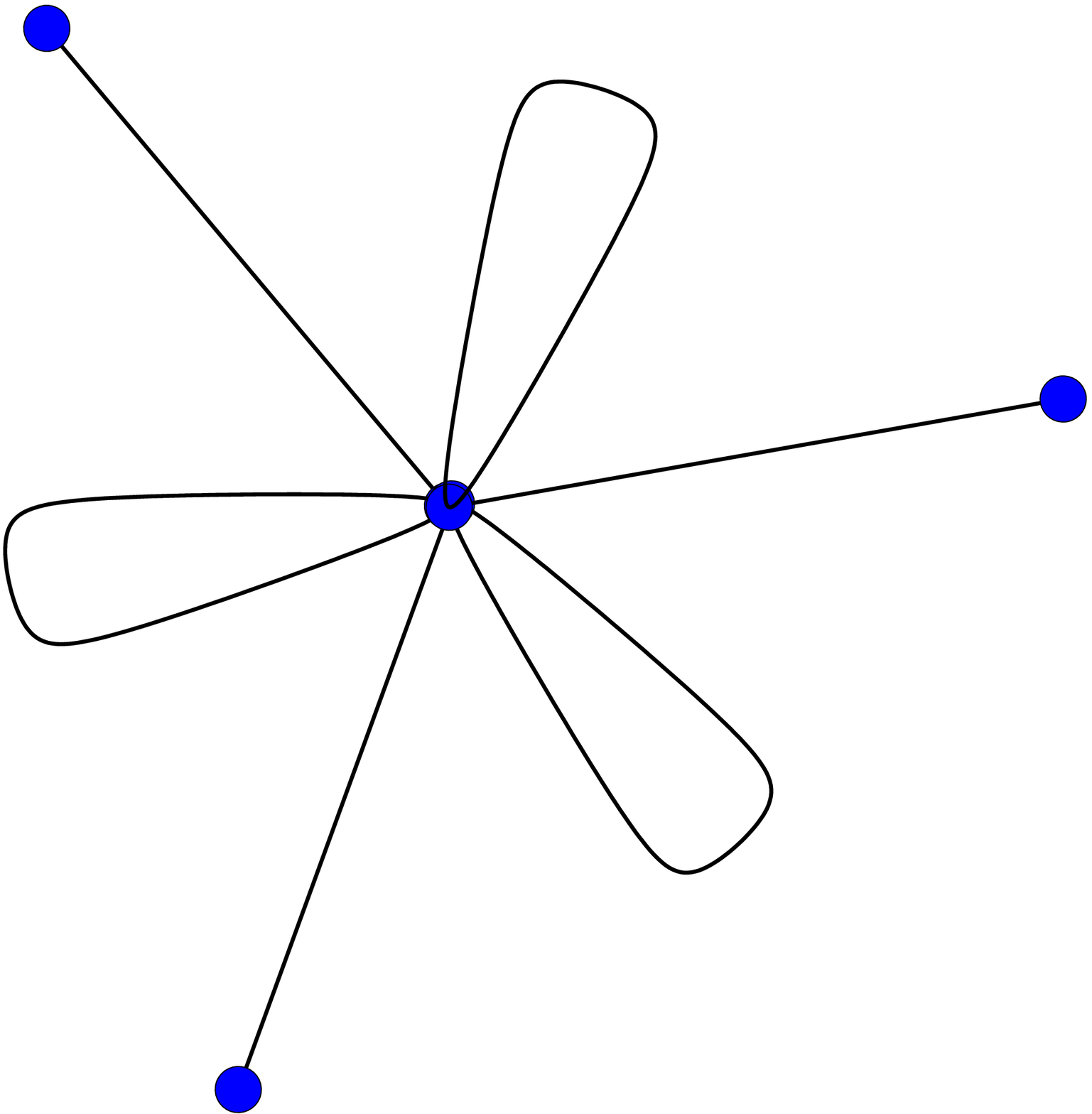}}}
  \ee
It has already been displayed in \cite{dizu1,difr2}, where also the $\zet_3$ 
fusion rules of the ordinary sectors were established by a different method..

These fusion rules provide the annulus multiplicities. Combining them with
the modular $S$-matrix of the $G_2$-theory yields 
three symmetry preserving boundary states,
  \be\begin{array}{ll}
  |00\rangle &= 3^{-1/4}\, \left(\, |00\Rangle
  + |11\Rangle + |02{,}+\Rangle + |02{,}-\Rangle\right) \,, \\[1.5mm]
  |02{,}+\rangle \!\!\!&= 3^{-1/4}\, \left(\, |00\Rangle
  + |11\Rangle + \eE^{2\pi\ii/3} |02{,}+\Rangle 
  + \eE^{-2\pi\ii/3} |02{,}-\Rangle \right)\,, \\[1.5mm]
  |02{,}-\rangle \!\!\!&= 3^{-1/4}\, \left(\, |00\Rangle
  + |11\Rangle + \eE^{-2\pi\ii/3} |02{,}+\Rangle 
  + \eE^{2\pi\ii/3} |02{,}-\Rangle \right)
  \end{array}\ee
as well as the single symmetry breaking boundary state
  \be |01\rangle = \Frac12\, 3^{1/4}\, (\sqrt3+\sqrt7)\, |00\Rangle
  + \Frac12\, 3^{1/4}\, (\sqrt3-\sqrt7)\, |11\Rangle \,.  \ee

We finally remark that the system of equations for the coefficients of 
the Ishibashi states is highly over-determined. We regard it as a
non-trivial check of our ideas that a solution exists at all.

\vskip2em
\noindent
{\small
{\sc Acknowledgement}:\\ We are grateful to J.\ B\"ockenhauer, D.E.\ Evans,
G.\ Felder, J.\ Fr\"ohlich and J.-B.\ Zuber for helpful discussions, and
to P.\ Bantay and B.\ Schellekens for a careful reading of the manuscript.
C.S.\ would like to thank the Schr\"odinger Institute for hospitality.}

 \vskip2.6em
\small
 \newcommand\wb{\,\linebreak[0]} \def\wB {$\,$\wb}
 \newcommand\Bi[1]    {\bibitem{#1}}
 \newcommand\J[5]   {{\sl #5}, {#1} {#2} ({#3}) {#4} }
 \newcommand\PhD[2]   {{\sl #2}, Ph.D.\ thesis (#1)}
 \newcommand\Prep[2]  {{\sl #2}, preprint {#1}}
 \newcommand\BOOK[4]  {{\em #1\/} ({#2}, {#3} {#4})}
 \newcommand\iNBO[6]  {{\sl #6}, in:\ {\em #1}, {#2}\ ({#3}, {#4} {#5})}
 \def\jf    {J.\ Fuchs}
 \def\adma  {Adv.\wb Math.}
 \def\aspm  {Adv.\wb Stu\-dies\wB in\wB Pure\wB Math.}
 \def\atmp  {Adv.\wb Theor.\wb Math.\wb Phys.}
 \def\comp  {Com\-mun.\wb Math.\wb Phys.}
 \def\ijmp  {Int.\wb J.\wb Mod.\wb Phys.\ A}
 \def\jgap  {J.\wb Geom.\wB and\wB Phys.}
 \def\joal  {J.\wB Al\-ge\-bra}
 \def\maan  {Math.\wb Annal.}
 \def\nupb  {Nucl.\wb Phys.\ B}
 \def\phlb  {Phys.\wb Lett.\ B}
 \def\phrl  {Phys.\wb Rev.\wb Lett.}
 \def\pnas  {Proc.\wb Natl.\wb Acad.\wb Sci.\wb USA}
 \def\rvmp  {Rev.\wb Math.\wb Phys.}
 \def\AP     {{Academic Press}}
 \def\NH     {{North Holland Publishing Company}}
 \def\SV     {{Sprin\-ger Ver\-lag}}
 \def\WS     {{World Scientific}}
 \def\Ad     {{Amsterdam}}
 \def\Be     {{Berlin}}
 \def\NY     {{New York}}
 \def\Si     {{Singapore}}

\end{document}